\documentclass{article}
\usepackage[preprint]{spconf}   
\usepackage{spconf,amsmath,graphicx,hyperref}
\usepackage{tikz}
\usetikzlibrary{arrows.meta,positioning,shapes.geometric,fit,backgrounds}
\usepackage{subcaption}
\usepackage{caption}
\usepackage{amssymb}   
\usetikzlibrary{positioning}
\usetikzlibrary{decorations.pathreplacing,positioning}
\usepackage{booktabs,multirow,tabularx,xcolor,colortbl}
\definecolor{headorange}{RGB}{230,115,50}
\definecolor{groupblue}{RGB}{206,222,246}
\definecolor{groupgreen}{RGB}{208,233,208}
\usepackage{float}
\usepackage{multirow}
\usepackage{bigdelim} 

\def\x{{\mathbf x}}

\title{AmbiDrop: Array-Agnostic Speech Enhancement Using Ambisonics Encoding And Dropout-Based Learning }
\name{Michael Tatarjitzky, Boaz Rafaely}
\address{School of Electrical and Computer Engineering, Ben Gurion University of the Negev, Beer-Sheva, Israel}
\begin{document}
\ninept
\topmargin=0mm 
\maketitle
\copyrightnotice{\parbox{\textwidth}{\footnotesize
\copyright~2026 IEEE. Personal use of this material is permitted. Permission from IEEE must be obtained for all other uses, in any current or future media, including reprinting/republishing this material for advertising or promotional purposes, creating new collective works, for resale or redistribution to servers or lists, or reuse of any copyrighted component of this work in other works.}}

\begin{abstract}
Multichannel speech enhancement leverages spatial cues to improve intelligibility and quality, but most learning-based methods rely on specific microphone array geometry, unable to account for geometry changes. To mitigate this limitation, current array-agnostic approaches employ large multi-geometry datasets but may still fail to generalize to unseen layouts.
We propose AmbiDrop (Ambisonics with Dropouts), an Ambisonics-based framework that encodes arbitrary array recordings into the spherical harmonics domain using Ambisonics Signal Matching (ASM). A deep neural network is trained on simulated Ambisonics data, combined with channel dropout for robustness against array-dependent encoding errors, therefore omitting the need for a diverse microphone array database.
Experiments show that while the baseline and proposed models perform similarly on the training arrays, the baseline degrades on unseen arrays. In contrast, AmbiDrop consistently improves SI-SDR, PESQ, and STOI, demonstrating strong generalization and practical potential for array-agnostic speech enhancement.
\end{abstract}
\section{Introduction}
\label{sec:intro}
Speech enhancement aims to improve the intelligibility and perceptual quality of speech in noisy and reverberant environments, with broad applications in teleconferencing, hearing aids, and human-machine interfaces. 
While single-channel deep neural networks (DNNs) have achieved impressive results \cite{osaughnessy2024speech}, multichannel methods can exploit spatial cues for further gains. Multichannel speech enhancement has progressed from classical beamforming such as delay-and-sum, MVDR, and super-directive methods \cite{VanTrees2002, Bitzer2001} to DNNs that jointly model spectral and spatial information \cite{10819706, huang2025advances}. However, most DNN-based methods assume a fixed microphone array geometry, thereby reducing robustness in real-world scenarios where array configurations vary across devices, form factors, and usage conditions.

To overcome the limitations of specific-array training, recent research has explored
array-agnostic deep learning approaches for speech enhancement. One prominent
strategy employs \emph{transform-average-concatenate} (TAC) layers
\cite{luo2020end, jukic2023flexible, yoshioka2022vararray}, which aggregate
information across channels in a permutation-invariant manner, enabling networks
to handle an arbitrary number of microphones. Another line of work applies
meta-learning \cite{mannanova2024meta}, allowing models to adapt to new array
geometries with limited data, but still requiring exposure to a wide
variety of array configurations during training. Alternative methods incorporate
more complex architectural designs—such as spatial transformers or
space-object cross-attention (SOCA) blocks \cite{taherian2022one}—which reduce
the need for multiple training arrays but often result in heavy models and
reduced robustness when facing unseen array layouts.
Overall, while these methods improve flexibility, they remain constrained: some
depend on large-scale training across diverse geometries, while others
generalize poorly to out-of-distribution arrays, highlighting the persistent
challenge of achieving truly array-agnostic enhancement.

In this work, we address the limitations of current multichannel speech enhancement methods by proposing a robust, truly array-agnostic approach that omits the need for an extensive multi-geometry training data while maintaining performance for a wide range of microphone arrays. The objective is to improve the generalization of DNN-based enhancement systems across diverse and unseen real-world array configurations.

To achieve this aim we introduce AmbiDrop, a DNN-based enhancement model trained on array-independent ideal Ambisonics representations~\cite{zotter2019ambisonics} to overcome the constraints of array-dependent training. Since accurate Ambisonics encoding may be challenging for most non-spherical arrays \cite{rafaely2015fundamentals}, with array-specific encoding errors, we incorporate input dropout during training to enhance robustness against missing or inaccurately encoded Ambisonics channels. This approach enables the representation of a broad range of previously unseen array configurations. While previous studies have used spherical harmonics (SH) as network inputs~\cite{pan2023hierarchical, pan2024innovative, pan2025enhancing}, they did not target the design of an explicitly array-agnostic network as proposed here. At inference, microphone signals are transformed into Ambisonics signals using Ambisonics Signal Matching (ASM) \cite{gayer2024ambisonics} prior to processing by the network. This preprocessing ensures that each channel has a consistent spatial interpretation, regardless of the original array geometry, thereby providing a stable, geometry-independent input representation that enables fast training without the need of extensive array data, leading to robust enhancement across diverse array configurations.

Finally, we evaluate the proposed framework on different unseen microphone arrays. The results show that while the baseline model with microphone signals as input performs well only on the arrays it was trained on, AmbiDrop maintains strong performance across all unseen array configurations, confirming its robustness and generalization capability. Moreover, informal subjective evaluations suggest that the enhanced signals exhibit only mild artifacts.
\begin{figure*}[t]
\centering

\begin{subfigure}[t]{\linewidth}
  \centering
  \small
  \begin{tikzpicture}[
    node distance=8mm and 7mm,
    >=Latex,
    every node/.style={align=center},
    block/.style={draw, rounded corners=2pt, line width=0.4pt, fill=black!8,
                  minimum width=2.cm, minimum height=0.7cm, inner sep=2.5pt},
    arrow/.style={-{Latex[length=2.5mm,width=1.7mm]}, line width=0.5pt}
  ]
    \node (amb) {Ideal\\Ambisonics};
    \node[block, right=of amb] (stft) {STFT};
    \node[block, right=of stft] (drop) {Input Channel-wise\\Dropout};
    \node[block, right=of drop] (net) {FT-JNF};
    \node[block, right=of net] (istft) {ISTFT};
    \node[right=of istft] (out) {Enhanced Speech};
    
    \draw[arrow] (amb) -- (stft);
    \draw[arrow] (stft) -- (drop);
    \draw[arrow] (drop) -- (net);
    \draw[arrow] (net) -- (istft);
    \draw[arrow] (istft) -- (out);
  \end{tikzpicture}
  \caption{Training stage}
  \label{fig:processing_chain_train}
\end{subfigure}

\vspace{4mm} 

\begin{subfigure}[t]{\linewidth}
  \centering
  \small
  \begin{tikzpicture}[
    node distance=8mm and 7mm,
    >=Latex,
    every node/.style={align=center},
    block/.style={draw, rounded corners=2pt, line width=0.4pt, fill=black!8,
                  minimum width=2.cm, minimum height=0.7cm, inner sep=2.5pt},
    arrow/.style={-{Latex[length=2.5mm,width=1.7mm]}, line width=0.5pt}
  ]
 \node (mic) {Microphone\\Signals};   
    \node [block, right=of mic](amb) {ASM};
    \node[block, right=of amb] (stft) {STFT};
    \node[block, right=of stft] (net) {FT-JNF};
    \node[block, right=of net] (istft) {ISTFT};
    \node[right=of istft] (out) {Enhanced Speech};
    
    \draw[arrow] (mic) -- (amb);
    \draw[arrow] (amb) -- (stft);
    \draw[arrow] (stft) -- (net);
    \draw[arrow] (net) -- (istft);
    \draw[arrow] (istft) -- (out);
    
  \end{tikzpicture}
  \caption{Inference stage}
  \label{fig:processing_chain_inference}
\end{subfigure}

\caption{Schematic diagram of the proposed model: (a) training and (b) inference.}
\label{fig:processing_chain}
\end{figure*}
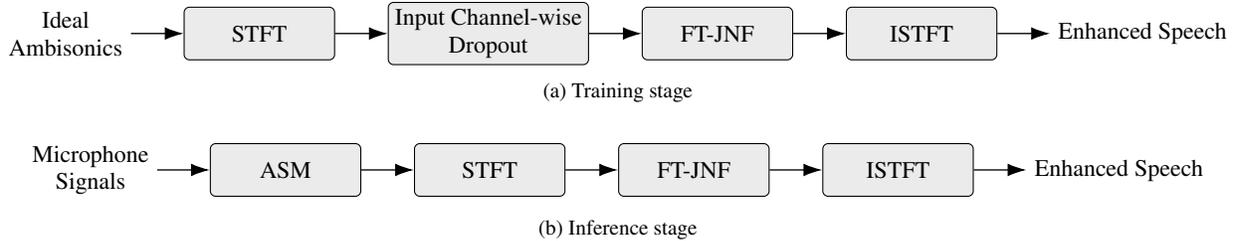
\section{Background}
\label{sec:background}
In this section, we present the signal model, the Ambisonics preprocessing framework, and the DNN architecture used in the remainder of the paper. The spherical coordinate system $(r, \theta, \phi)$ represents the radius, elevation, and azimuth respectively, while the wave number is defined as $k = \frac{2\pi}{c} f$, where $c$ is the speed of sound and $f$ is the frequency. 

\subsection{Array Signal Model}
\label{ssec:signal_model}
Consider $M$ omnidirectional microphones of an arbitrary array positioned in a sound field, with each microphone at coordinates $(r_m, \theta_m, \phi_m)$, $\forall\, 1 \leq m \leq M$. 
Assume the sound field is composed of $Q$ plane waves with directions of arrival (DOA) $(\theta_q, \phi_q)$, $\forall\, 1 \leq q \leq Q$, denoted as $\Omega_Q$. The plane waves represent direct sound from sources or their reflections from enclosure boundaries.
The array steering matrix is defined as $\mathbf{V}(k)$ with dimensions $M \times Q$ where each element $[\mathbf{V}(k)]_{m,q}$ is the frequency-dependent transfer function between the $q$-th plane wave and the $m$-th microphone.
The signal recorded at the microphones can then be expressed as:
\begin{equation}
     \mathbf{x}(k) = \mathbf{V}(k) \, \mathbf{s}(k) + \mathbf{n}(k),
\end{equation} 
Here, $\mathbf{x}(k)$ is a vector of length $M$, where each element $x_m(k)$ corresponds to the $m$-th microphone signal, $\forall\, 1 \leq m \leq M$. 
Similarly, $\mathbf{s}(k)=[s_1(k),...,s_Q(k)]$ is a vector of length $Q$, where each element $s_q(k)$ denotes the amplitude at the origin of the plane wave arriving from the $q$-th DOA, $\forall\, 1 \leq q \leq Q$.
The term $\mathbf{n}(k)$ represents additive noise at the microphones, assumed to be i.i.d.\ across microphones and uncorrelated with the source signals $\mathbf{s}(k)$. 

\subsection{Ambisonics Encoding}
\label{ssec:ASM_preprocessing}
The Ambisonics signal vector can be obtained from the source signals vector as (see Eq.~2.43 in~\cite{rafaely2015fundamentals}):
\begin{equation}
    \mathbf{a}_{nm}(k) = \mathbf{Y}^H_{\Omega_Q} \, \mathbf{s}(k),
\label{eq:amb}
\end{equation}
Here $\mathbf{a}_{nm}(k) = [a_{00}(k), \dots, a_{N_a N_a}(k)]^T$ denotes the Ambisonics signals up to order $N_a$, with length of $(N_a + 1)^2$. The matrix $\mathbf{Y}^H_{\Omega_Q} = [\mathbf{y}_{00}, \dots, \mathbf{y}_{N_a N_a}]$ is the SH matrix of dimensions $Q \times (N_a + 1)^2$, where each vector $\mathbf{y}_{nm} = [y_{nm}(\theta_1, \phi_1), \dots, y_{nm}(\theta_Q, \phi_Q)]^T$ is of length $Q$ containing the SH basis functions of order $n$ and degree $m$ evaluated at the $Q$ DOAs $(\theta_q, \phi_q)$, for all $1 \leq q \leq Q$ with $0 \leq n \leq N_a$ and $-n \leq m \leq n$.

Ambisonics Signal Matching (ASM) \cite{gayer2024ambisonics} is a method for encoding Ambisonics signals from arbitrary microphone array configurations using a linear mapping from the microphone signals to the $(n,m)$-th Ambisonics coefficient:
\begin{equation}
    \hat{a}_{nm}(k) = \mathbf{c}_{nm}^H \mathbf{x}(k),
    \label{eq:amb_est}
\end{equation}
where $0 \leq n \leq N_a$ and $-n \leq m \leq n$.  
Here, $\mathbf{c}_{nm}$ is an $M \times 1$ vector of filter coefficients designed to minimize the normalized mean square error (NMSE) between the true and estimated Ambisonics coefficients:
\begin{equation}
\varepsilon_{\mathrm{Amb}} = 
\frac{E\left[ \lVert \hat{a}_{nm}(k) - a_{nm}(k) \rVert_2^2 \right]}
     {E\left[ \lVert a_{nm}(k) \rVert_2^2 \right]},
\label{eq:amb_error}
\end{equation}
where $E[\cdot]$ denotes the expectation operator and $\lVert \cdot \rVert_2$ is the $\ell_2$ vector norm.  
Assuming the noise vector $\mathbf{n}(k)$ is spatially white with power $\sigma_n^2$ and the sound field is diffuse—modeled as $Q$ uncorrelated plane waves such that $\mathbf{R}_s(k) = \sigma_s^2 \mathbf{I}$—the optimal filter coefficients are obtained by solving \eqref{eq:amb_error}, yielding:
\begin{equation}
\mathbf{c}_{nm}^{\mathrm{opt}}(k) =
\left( \mathbf{V}(k)\mathbf{V}^H(k) + \frac{\sigma_n^2}{\sigma_s^2} \mathbf{I} \right)^{-1} 
\mathbf{V}(k) \mathbf{y}_{nm},
\label{eq:c_opt}
\end{equation}
Substituting \eqref{eq:c_opt} into \eqref{eq:amb_est} produces the estimated Ambisonics signals. To encode all Ambisonics channels, the following condition must hold \cite{rafaely2015fundamentals}:
\begin{equation}
    (N_a + 1)^2 \leq M.
\end{equation}
It was shown in \cite{gayer2024ambisonics} that some Ambisonics channels may not be accurately encoded, depending on array geometry and the array steering matrix.

\subsection{DNN Architecture}
\label{ssec:DNN_architecture}
In this paper, we employ a speech enhancement network based on FT-JNF~\cite{tesch2022insights}. 
We consider a speech enhancement task in which the goal is to extract the target speaker from a noisy and reverberant recording.

Applying the short-time Fourier transform (STFT) to $x_m(t)$, the signal in the $m$-th microphone, yields the time--frequency representation $X_m(t,f)$ where $(t,f)$ is a time-frequency bin.
In the presence of noise, the noisy microphone signal $X_m(t,f)$ can be modeled as:
\begin{equation}
    X_m(t,f) = Y_m(t,f) + N_m(t,f),
\end{equation}
where $Y_m(t,f)$ includes the clean speech of the target speaker, $S(t,f)$, as recorded by the $m$-th microphone and $N_m(t,f)$ represents the noise component captured by the $m$-th microphone, which may include interfering speakers, reverberation and background noise. The multichannel microphone array signal is represented as $\mathbf{X}(t,f) = [X_1(t,f),...,X_M(t,f)]$, a tensor of dimensions $M\times T\times F$.

The network estimates a complex ideal ratio mask (cIRM) from multichannel inputs by jointly exploiting spatial, spectral, and temporal information. 
The enhanced signal is obtained by applying the estimated mask \( M(t,f) \) to a reference noisy channel \( X_1(t,f) \), here chosen as the first microphone channel:
\begin{equation}
    \hat{S}(t,f) = M(t,f) \cdot X_1(t,f)
    \label{eq:net_output}
\end{equation}
The network consists of three layers, two bidirectional long short-term memory (BLSTM) layers, followed by a linear layer. 

\newcommand{\arraycircle}{
  \draw[dashed] (0,0) circle (1);
  \draw[black!20] (-1.1,0) -- (1.1,0);
  \draw[black!20] (0,-1.1) -- (0,1.1);
}

\newcommand{\plotmics}[1]{%
  \foreach \x/\y in {#1} { \filldraw[black] (\x,\y) circle (1.7pt); }%
}

\newcommand{\pmic}[2]{\filldraw[black] ({#2*cos(#1)},{#2*sin(#1)}) circle (1.7pt);}

\newcommand{\arraysubfig}[2]{%
\begin{subfigure}[t]{0.31\linewidth}
  \centering
  \begin{tikzpicture}[scale=1]
    \arraycircle
    #2
    \draw[black!50] (-1.1,-1.1) rectangle (1.1,1.1); 
  \end{tikzpicture}
  \caption{#1}
\end{subfigure}
}
\begin{figure*}[t]
\centering
\begin{minipage}[t]{0.48\textwidth}
\centering
\arraysubfig{full circle $r=0.1m$}{%
  \pmic{0}{1} \pmic{72}{1} \pmic{144}{1} \pmic{216}{1} \pmic{288}{1}
}
\arraysubfig{semi circle $r=0.05m$}{%
  \fill (0,-0.5) circle (2pt);
  \fill (0.354,-0.354) circle (2pt);
  \fill (0.5,0) circle (2pt);
  \fill (0.354,0.354) circle (2pt);
  \fill (0,0.5) circle (2pt);
}
\arraysubfig{ULA (Y-axis)}{%
  \plotmics{0/-1, 0/-0.5, 0/0, 0/0.5, 0/1}
}

\vspace{2mm}

\arraysubfig{X shape}{%
  \plotmics{0/0}
  \pmic{45}{1} \pmic{135}{1} \pmic{225}{1} \pmic{315}{1}
}
\arraysubfig{random1}{%
  \fill (0.454,-0.096) circle (2pt);
  \fill (-0.363,-0.354) circle (2pt);
  \fill (-0.167,0.299) circle (2pt);
  \fill (0.452,0.478) circle (2pt);
  \fill (0.740,0.362) circle (2pt);
}
\arraysubfig{random2}{%
  \fill (0.093,0.805) circle (2pt);
  \fill (0.726,-0.539) circle (2pt);
  \fill (-0.4,0.793) circle (2pt);
  \fill (0.496,0.402) circle (2pt);
  \fill (-0.187,0.327) circle (2pt);
}

\caption{Training arrays ($M=5$, radius $=0.10\,m$).}
\label{fig:train_arrays}
\end{minipage}
\hfill
\begin{minipage}[t]{0.48\textwidth}
\centering
\arraysubfig{full circle $r=0.05m$}{%
  \pmic{0}{0.5} \pmic{72}{0.5} \pmic{144}{0.5} \pmic{216}{0.5} \pmic{288}{0.5}
}
\arraysubfig{semi circle $r=0.1m$}{%
  \fill (0,-1) circle (2pt);
  \fill (0.707,-0.707) circle (2pt);
  \fill (1,0) circle (2pt);
  \fill (0.707,0.707) circle (2pt);
  \fill (0,1) circle (2pt);
}
\arraysubfig{ULA (X-axis)}{%
  \plotmics{-1/0, -0.5/0, 0/0, 0.5/0, 1/0}
}

\vspace{2mm}

\arraysubfig{plus shape}{%
  \plotmics{0/0}
  \plotmics{0/1, 0/-1, 1/0, -1/0}
}
\arraysubfig{random3}{%
  \fill (0.222,0.542) circle (2pt);
  \fill (-0.916,0.012) circle (2pt);
  \fill (-0.329,0.023) circle (2pt);
  \fill (-0.082,0.411) circle (2pt);
  \fill (0.610,-0.247) circle (2pt);
}
\arraysubfig{random4}{%
  \fill (0.906,-0.037) circle (2pt);
  \fill (0.757,-0.342) circle (2pt);
  \fill (-0.825,-0.061) circle (2pt);
  \fill (-0.609,-0.635) circle (2pt);
  \fill (0.398,0.087) circle (2pt);
}

\caption{Test arrays ($M=5$, radius $=0.10\,m$).}
\label{fig:test_arrays}
\end{minipage}
\end{figure*}

\section{Proposed Method}
\label{sec:method}

We propose a novel method for array-agnostic speech enhancement that eliminates the need for training on multiple microphone array geometries. Our approach leverages Ambisonics representations as network input, which, as shown in \eqref{eq:amb}, depend solely on the sources and reverberant sound field rather than a specific microphone array. However, such an idealized representation of the sound field cannot be directly obtained from real recordings, due to the typically irregular arrangements of the microphones. Nevertheless, the proposed training with dropouts accounts for this aspect, as explained below.

\subsection{Training Stage}
During training (Fig.~\ref{fig:processing_chain_train}), the network is provided exclusively with simulated Ambisonics signals up to order $N_a$, ensuring complete independence from array geometry. The input $\mathbf{a}_{nm}(t)$ is transformed via STFT to $\mathbf{A}_{nm}(t,f)$, which is then passed through a channel-wise dropout layer before entering the FT-JNF module.  
The omnidirectional harmonic $A_{00}(t,f)$ serves as the reference channel due to its directional uniformity. The network estimates a complex ideal ratio mask (cIRM), which is applied to the reference channel as in~\eqref{eq:net_output} to produce the STFT representation of the enhanced signal. The time-domain enhanced signal is then reconstructed via inverse STFT. Training is optimized using the SI-SDR loss~\cite{le2019sdr}.

\subsection{Inference Stage}
At inference (Fig.~\ref{fig:processing_chain_inference}), multichannel microphone signals from arbitrary arrays are transformed into Ambisonics signals using ASM \cite{gayer2024ambisonics}. These are then processed through the same chain as in training, except the dropout layer is bypassed, allowing the model to exploit all available input channels.

\subsection{Insight Into The Dropout Process}
ASM encoding may yield imperfect estimates of certain Ambisonics channels, depending on the array geometry \cite{gayer2024ambisonics}. In particular, poorly estimated channels tend to have low amplitude, approaching zero in cases that estimation is not possible (see Section 4 in \cite{gayer2024ambisonics}). This behavior is analogous to dropout in neural networks, where some input channels are randomly suppressed. By incorporating channel-wise dropout during training, we implicitly simulate this phenomenon, thereby enhancing robustness against imperfect Ambisonics encoding in real-world data. A preliminary ablation study showed that removing channel-wise dropout degraded generalization, highlighting its importance.

\subsection{Insight Into The Impact of Ambisonics Input}
Training DNNs directly on microphone array signals requires exposure to multiple array geometries, since spatial relations across channels vary with array layout. In contrast, Ambisonics signals provide a fixed channel count with a consistent spatial relations across channels. This stability simplifies the training process, allowing the network to learn spatial cues from a uniform representation and thereby generalize to unseen arrays without relying on a large multi-geometry dataset.
\begin{table*}[!t]
\centering
\fontsize{9}{13}\selectfont
\setlength{\tabcolsep}{5pt}
\begin{tabular}{|c|c|cc|cc|cc|}
\hline
\multirow{2}{*}{\textbf{Dataset}}& \multirow{2}{*}{\textbf{Method}}&
\multicolumn{2}{c|}{\textbf{SI-SDR} (dB) {$\uparrow$}} &
\multicolumn{2}{c|}{\textbf{PESQ} {$\uparrow$}} &
\multicolumn{2}{c|}{\textbf{STOI} {$\uparrow$}} \\ \cline{3-8}
 & & Noisy & Enhanced & Noisy & Enhanced & Noisy & Enhanced \\ \hline

\multirow{2}{*}{Training Arrays}& Baseline & -7.2 & \textbf{5.6}& 1.18 & 1.73& 0.58 & \textbf{0.84}\\
 & Proposed & -7.7 & 3.9& 1.20 & \textbf{1.84}& 0.58 & 0.83\\ \hline

\multirow{2}{*}{Test Arrays}& Baseline & -7.2 & -7.4& 1.18 & 1.32& 0.58 & 0.64\\
 & Proposed & -6.6 & \textbf{5.4}& 1.14 & \textbf{1.90}& 0.58 & \textbf{0.86}\\ \hline

\multirow{2}{*}{AR glasses}  
 & Baseline & -3.0& -40.1& 1.22& 1.34& 0.68& 0.28\\
 & Proposed & -9.0& \textbf{-2.0}& 1.18& \textbf{1.59}& 0.59& \textbf{0.75}\\ \hline

\end{tabular}
\caption{Mean performance on train, test, and AR glasses arrays. Each metric (SI-SDR, PESQ, STOI) is reported for noisy and enhanced signals under the baseline and proposed models.}
\label{tab:train_test_ar}
\end{table*}

\section{Experiment}
\label{sec:experiment}

In this section, we evaluate AmbiDrop (Section~\ref{sec:method}) in a multi-speaker scenario, where the task is to extract one target speaker from a mixture of six speakers. For comparison, we use the FT-JNF network with microphone signals input as an array-dependent baseline.

\subsection{Setup}
\label{ssec:Setup}

We generated multiple scenes using the image method \cite{allen1979image} implemented in MATLAB. The room and source configurations followed the setup in \cite{tesch2022insights} (section~4.A), with the only difference being the microphone array designs. All simulated arrays were planar (2D), with five microphones placed within a circle of radius 0.1\,m, and with their steering vectors modeled under free-field conditions. In addition, we incorporated a real far-field array steering vector from head-worn AR glasses provided in the EasyCom dataset \cite{donley2021easycom}.

For the \textbf{baseline model}, training and validation data were generated using six different arrays with diverse geometries (see Fig. \ref{fig:train_arrays}), to encourage generalization across array types. The clean reference signal was defined as the direct path of the target speaker, captured by the reference microphone located closest to the target source, and additive noise was introduced at an SNR of 30~dB.

For \textbf{AmbiDrop}, the training and validation datasets consisted of ideal simulated Ambisonics signals up to order $N_a = 2$, generated directly from the source signals and reverberation components constructed using the image method. The clean reference signal was defined as the omnidirectional spherical harmonic coefficient $a_{00}(t)$ of the direct path of the target speaker. Since this study focuses exclusively on 2D arrays, we restricted the Ambisonics representation to 2D Ambisonics (circular harmonics), which is the subset of harmonics with the highest azimuthal resolution, i.e., $m = \pm n$ (see Chapter~1.2 in \cite{rafaely2015fundamentals}), resulting in a five-channel input.

During inference, both models were evaluated on three datasets containing the same speech mixtures but with different array geometries. The first consisted of the same array geometries used for training the baseline model, while the second included six previously unseen arrays with distinct geometrical layouts (see Fig. \ref{fig:test_arrays}). The third dataset incorporated a real-world AR glasses array, using the first five microphones of the array (see Fig. 2 in \cite{donley2021easycom}).
For AmbiDrop, microphone signals were first transformed into Ambisonics representations via ASM (Section~\ref{ssec:ASM_preprocessing}) assuming an SNR of 30 dB.

We used clean speech signals from the WSJ0 dataset \cite{garofolo2007wsj0}, sampled at 16~kHz. The baseline and AmbiDrop used 6000 training and 1002 validation samples, with 300 test samples per array and no dataset overlap. 

\subsection{Methodology}
\label{ssec:methodology}
For training and validation, 6-second segments of clean and noisy signals were extracted and aligned to the clean speech onset. The STFT used a 32-ms Hamming window with 50\% overlap. 

For the DNN (see Section \ref{ssec:DNN_architecture}) in both of the models, the BLSTM layers were configured with 256 and 128 units, respectively. The output of the final linear layer had size 2, corresponding to the real and imaginary components of the estimated mask. In \textbf{AmbiDrop} (see Section \ref{sec:method}), we applied channel-wise dropout, randomly dropping up to three channels with probability 0.4. Both models were trained with a batch size of 8, weight decay of $10^{-5}$, and the Adam optimizer with a learning rate of 0.001. Each network was trained for 250 epochs, and the model corresponding to the epoch with the lowest validation loss was selected.

Model performance was evaluated using three objective metrics: scale-invariant signal-to-distortion ratio (SI-SDR), perceptual evaluation of speech quality (PESQ), and short-time objective intelligibility (STOI).

\subsection{Results}
\label{ssec:results}
As described in Section \ref{ssec:Setup}, we conducted three experiments, and the results are summarized in table \ref{tab:train_test_ar}.
In the first experiment, the arrays belonged to the baseline model’s training set but were unseen for AmbiDrop. Both approaches achieved comparable performance across all evaluation metrics, with the baseline showing a slight advantage in SI-SDR (1.7~dB), reflecting its specialization on the seen geometries. AmbiDrop, despite never being exposed to these arrays during training, maintained similar performance, demonstrating strong generalization. Notably, while AmbiDrop achieved slightly lower SI-SDR, it outperformed the baseline in PESQ. This suggests that although both models were trained with SI-SDR as the optimization objective, AmbiDrop was able to preserve perceptual quality more effectively, leading to improved PESQ as a beneficial side effect.

In the second and third experiments, the models were tested on unseen configurations: six simulated layouts and a real AR glasses array. In these settings, the baseline suffered substantial performance degradation, whereas AmbiDrop maintained strong results across all metrics, with only a moderate drop in performance for the AR glasses array. Possible cause for the latter could be the different array configuration which includes scattering from the head and microphone distribution that is not strictly in two dimensions. These findings highlight AmbiDrop’s superior robustness and ability to generalize effectively to both synthetic and real-world arrays, confirming its practicality for diverse deployment scenarios.

\section{Conclusions}
\label{sec:conclusions}
This work introduced AmbiDrop, a novel Ambisonics-based framework for array-agnostic speech enhancement. By leveraging Ambisonics encoding as a geometry-independent representation and incorporating channel-wise dropout to mitigate encoding errors, the method eliminates the need for training across multiple microphone geometries.

Through experiments on simulated and real-world arrays, we demonstrated that the proposed approach maintains competitive performance on arrays that are seen for the baseline, while substantially outperforming the baseline model when evaluated on unseen arrays. Notably, AmbiDrop generalizes effectively even to a real-world AR glasses array, confirming its robustness to both synthetic and practical array configurations. While experiments included only arrays with 5 microphones, the Ambisonics representation could potentially generalize to other microphone counts, an aspect which is beyond the scope of this study.
Overall, AmbiDrop provides a scalable and practical solution for multichannel speech enhancement across diverse recording devices.

Future work is proposed to focus on evaluating the AmbiDrop framework with alternative DNN architectures and dropout processes, and validate its performance on real-world datasets. 

\vfill
\pagebreak

\bibliographystyle{IEEEbib}
\bibliography{AmbiDrop}

\end{document}